# On the Peculiarities of Design: An Engineering Perspective

*Mouhamed Abdulla and Yousef R. Shayan*
Department of Electrical and Computer Engineering, Concordia University, Montréal, Québec, Canada
Email: {m_abdull, yshayan} @ ece.concordia.ca

***Abstract** –It is a fact of our existence, that no matter where we are, we most often find ourselves either hearing, seeing, talking, or even engaged in design related activities. Despite this reality, the notion of "design", and in particular "engineering design", is often ambiguous, and at times obscure. Thus, the transfer of knowledge of this crucial topic to engineering students engaged in practical hands-on learning or analytical research is usually perplexing to accomplish. In light of this, it becomes worthwhile to dissect and reflect on the abstraction of the design process in engineering. In fact, the aim of this article is to investigate the facets of applied design, and elaborate on its diversity, complexity and elements. Eventually, by concretizing this subject matter, we hope to slightly assist engineering students in alleviating some of the vagueness associated with the principle of design, and enhance their technical skillset during innovative conceptions.*

## 1. INTRODUCTION

In general, the widely established and practiced approach for academic learning is based on attending lectures, studying, and taking exams. Although this acquisition procedure does contain some merit, one of its major drawbacks is that inevitably most students will quickly forget the material once the course is over. An improved model would rather inclusively focus on adequate transfer of knowledge and technical skills; on proper understanding and competence assessment; and more importantly on active learning through hands-on practice. In fact, a balance of these criteria should ensure a higher quality of learning because as a byproduct, students would retain the subject more so than otherwise; they would be professionally prepared for the workforce; and would have the necessary prerequisites and competence to even innovate. Clearly, the innovation and discovery aspect requires at root a certain passion and enjoyment of the field, and this cannot necessarily be acquired; however, it can always be facilitated through well-trained and dedicated instructors.

Admittedly, when pondering on today's scientific progress, it becomes very hard not to be intellectually overwhelmed by it. Despite the outcome of this incredible evolution, which in essence resulted in numerous specializations, fields and subfields, in the larger scheme of things, they all share common attributes. Essentially, in all branches of pure and applied sciences, in-field training is indispensible. In fact, what sets them apart are the distinct objectives that each carry. For instance, the aim of pure sciences is to advance our understanding of nature through the search of cognitive knowledge in the form of new laws and universal truth. On the other hand, the motive for engineering is the technical uncovering of a useful, practical and efficient system, designed specifically to serve and assist the needs, desires and aspirations of a segment of people or society at large [1].

Besides, the primal role of professional engineers can further be defined as a rational group able to collectively transform or improve a practical idea into a system through the Conceive, Design, Implement, and Operate (CDIO) steps. As a matter of fact, historically speaking, CDIO$^{TM}$ is a term coined by MIT's Department of Aeronautics and Astronautics in the late 1990s [2]. In short, the goal of CDIO is to reform engineering education by complementing the traditional pedagogical model with practical hands-on experience in order to bridge and hopefully shorten the gap between academia and industry. In other words, the idea of the CDIO initiative is located at the intersection between the notion of active learning and the essential sciences of applied engineering.

Certainly, as noted above, design plays a pivotal role in the system lifecycle of CDIO. And as the global reach of this newly ameliorated scholastic model gains momentum, it becomes critical and natural to analyze with some depth the different approaches, levels, and specifications of the design process. Thus, in this paper, we will explore the philosophical rendering of engineering design for the purpose of identifying solid and effective descriptions for students. We also intend to examine the different varieties of design and analyze the interconnectedness of the methods. Finally, we will highlight the importance to bring all aspects together for a comprehensive design methodology. Despite the philosophical nature of the topic, our aim is to make the flow of this treatment, as much as possible, systematic and coherent.

## 2. UNDERSTANDING DESIGN

Every field of study is expressed by a particular interpretation language. In applied sciences, this interpretation is by and large dependent on engineering design. In fact, whether we are aware of it or not, the frequency of our interaction with design is relatively high. Although this is true, there still remains a degree of vagueness and an associated difficulty to clearly identify what design is all about.

A fundamental cause might be the lack of an explicit characterization of the term due perhaps to an unreachable consensus. And probably this is the case because of the varied nature of the expression, which by necessity resulted in a multitude of likely definitions. Also, what further adds to the following intricacy is that these definitions are not firm; in fact they are open for interpretation from either a scientific or a philosophical perspective.

Without a doubt, over the past decennium, the body of knowledge that encapsulates engineering design has been treated and expanded by thinkers, authors, and academics. And naturally, their findings and thoughts have been published in journals, survey papers, and even books (e.g. [3] and [4], among many other great contributions). Nonetheless, without trivializing these important and carefully crafted analyses, there still remain vast areas and angles that require a relook with a different perspective. Consequently, the objective of this treatment is to slightly move in this direction and attempt to explicitly breakdown and seek a practical and simple understanding of design from an engineering standpoint in order to gradually advance its teaching and learning.

## 3. MANAGEMENT OF DESIGN

Virtually all intelligent beings organize *a priori* a sequence of events that they set to follow. In other words, to achieve a particular purpose no matter the degree of difficulty, a certain phase of planning is required. And, the rational for planning is to have a clear framework and an agenda for the sequence of instances and their respective priorities. Further, these priorities and guidelines could take one of two forms, either for self or group management.

This sort of organization not only promotes a directed and focused goal, but also helps in assessing the different gradual milestones and the overall progressiveness of the project; which is largely a function of productivity. Thus, in this context, we can associate the discrete concept of design as:

*the intended action of organizing, planning and executing a task to achieve a particular purpose.*

## 4. CREATIVITY OF DESIGN

Perhaps less explicative, design can also be seen from an artistic outlook. Although this sense is usually inherent, yet in this situation it is still possible to get some training to enhance a subset of the needed qualifications. The same is also true for creativity. In other words, there are no direct specifications to fully characterize it. But broadly speaking, creativity can be helped in different ways. For instance, a designer can bring together two separate ideas to form a novel notion all together. Incidentally, the widely celebrated inventor, Steve Jobs, is known to have said that: *Creativity is just connecting things*. However, this thought process requires alertness and agility in order to identify and elegantly connect the different dots. Thus, at base, engineering creativity demands commonsense coupled with a strong insight of basic fundamentals.

Moreover, it is imperative to stress out that the goodness of an idea or a vision does not necessarily depend on its perplexity, but rather on its novelty, practicality, and its reception within the targeted community. In fact, a number of brilliantly discovered ideas in engineering and otherwise were quite simple. The so-called "Alamouti code" is a good and relatively recent example in wireless communications; where interestingly enough the title of this well-known paper is called "A *simple* transmit diversity technique for wireless communications" [5].

Another good way to improve creativity is to develop better observation skills of the surrounding and environment of interest, and then detect possible ways to further elevate the usefulness of an object or system. Of course, observation should not be limited to manmade artifacts, but could very well extend to natural phenomena. In fact, a new trend in applied sciences commonly known as "bio-inspired system design" has recently emerged. The essential drive of this paradigm is to find solutions to complicated engineering problems through the analysis and scrutiny of natural realities.

A fascinating example that illustrates these points is the famed case of Robert Kearns, Ph.D. [6], which in 2008 was adapted into a Hollywood movie (Flash of Genius). The story is mostly known for the struggle faced regarding the patent infringement by Ford and Chrysler of his "intermittent windshield wiper system". Legal issues aside, from an innovative and design perspective we notice the following gradual progression:

1. *Kearns used an already available system:* the original wipers in his 1960s Ford Galaxie.

2. *After interaction with the system over different conditions, he identifies an important flaw:* the wipers continuously move with only two settings; steady and heavy rain.

3. *He then notices the sophistication of nature:* while looking at the mirror he is intrigued by the way the eyelids work.

4. *He then identifies a potential liaison between events:* his dissatisfaction of the wipers could be solved by drawing a parallel with his enthrallment of the eyelids; since the eyes blink every couple of seconds as opposed to continuously.

5. *He transforms his idea into a functional system:* because of his engineering background and his understanding of circuit analysis using mainly transistors, capacitors, and variable resistors he is able to conceive, design, implement, and operate his vision.

What is remarkable is that using intelligent observation, common sense, analogy from nature, and the ability to connect ideas, Kearns was able to redesign a system by adding a simple yet practical feature to ameliorate its functionality. Although in the above case we nicely see how the thought process is used to conceive a genuine idea, it is worthy to note that the inspiration for innovation is for the most part uncontrollable. In other words, it will emerge intrinsically once it is ready. Thus, we ought to remark that original creativity should be facilitated rather than intentionally forced.

Additionally, an inventor should remain pragmatic and to some extent objective about the actual capability of a vision. Namely, one should be careful not to get too connected to an idea and stray from what others think of the invention. After all, the opinion of the client is largely part of the success metric for a concept. Thus, market research feedback is fundamental in adjusting and perfecting a system.

## 5. EXECUTION OF DESIGN

In the above sections, we separately identified design from a management and creativity point of views. However, to have a more balanced and inclusive understanding for the notion of design, it becomes important to merge these distinct features. In fact, for good design, we need an exceptional idea. However, an idea by itself is only part of the formula. The execution of the idea is as important, or at times more so, than the vision itself. Thomas Edison, the famous inventor, once said that: *Genius is one percent inspiration and ninety-nine percent perspiration.* Thus, although the "inspiration" is what initially triggers the first step of innovative design, in the long run, the "planning" and the "doing" of an idea through hard work seems substantially vital. And, this is quite natural because the concretization and transformation of an idea into a system enables factual testing and verification of the prototype. Moreover, design is an iterative process, where improvements are continuously applied as flaws are identified through hands-on experimentation until satisfaction. So no matter the adopted definition, design fundamentally consists on the thought process required in doing a precise deed.

## 6. ROLE OF DESIGN IN ENGINEERING

Design can take different forms in different scenarios and disciplines. And, in particular, it is widely treated in many facets of engineering. But how exactly are they related to each other? In other words, how would each as a standalone entity influence or complement the other? This will form the essence of our discussion here.

Engineering, at its core, focuses on all the technical aspects essential in making a useful system operate. It usually requires a solid foundation in the principles of a subject and an exceptional talent in manipulating the different analytic and physical tools or techniques. At the same time, over a range of circumstances, there are certain limitations that must be respected for rigorous engineering. In fact, these constraints come in different forms. Broadly speaking, they fall under one of two categories: either fundamental limits of a particular engineering field (e.g. Shannon's capacity in information theory) or practical restrictions quantified by a specific metric indicator (e.g. systems performance, efficiency, feasibility, quality of service, etc). To reconcile these restrictions, whether fundamental or practical, we will inevitably and inescapably require the essence of design in order to converge to a solution.

Indeed, we should stress that in design we generally refer to the outcome as a "solution" as opposed to an "answer" [7]. In fact, this is true because design will not produce an absolute right or wrong reaction; it will only give possible explanations or workarounds to a particular problem. In other words, if an engineering problem is given under near idealistic or oversimplified specifications, such as: well-defined, well-behaved and perhaps deterministic, then it is normally possible to provide a firm answer using elementary methods. However, accurate real-life engineering, which is relatively multifaceted, is founded on an approximate truth influenced significantly by tradeoffs.

In other words, design is an "open-ended" task. To illustrate this concept, say we have $m$-independent engineering teams working on the same exact design project. In this event, it can be anticipated, with a relatively high likelihood, that the teams will actually generate $m$-distinct design solutions. Clearly, the optimum solution to be picked will be the best among the $m$-results. And this obviously creates a challenge for properly identifying the criteria and performance indicators to assess the goodness of a design, which is not necessarily a trivial endeavor. Nonetheless, it will without a doubt be a function of the clients' requirements.

Meanwhile, we should recognize that factual engineering demands design; whereas the reverse is not necessarily required. For example, a domestic designer does not necessarily require engineering expertise; while a wireless engineer ought to rely on system design in order to deal with often contradicting requirements and ensure market penetration. That is to say that the relation between "engineering" and "design" is unidirectional. To further illustrate this point of view, in Fig. 1 we coherently illustrate a possible and simple interpretation for engineering design, where the exclusive contribution of each term is emphasized.

## 7. ENTANGLEMENT OF DESIGN

At this level, we could demonstrate the application of design via a system engineering problem, where the challenge is to identify, address, and reconcile the various competing requirements. For instance, in wireless engineering, the broad objective is to ensure that the transmitted signal remains reliable as it travels from source to destination. In fact, we could address this problem progressively from simple to complex by gradually adding natural and manmade limitations [8]. Further, as these limitations are introduced, the task of adequately adjusting the needed requirements, or appropriately designing the full system, becomes significantly complicated. As an example, some of the entanglements often faced by a link-layer system designer are shown in Fig. 2.

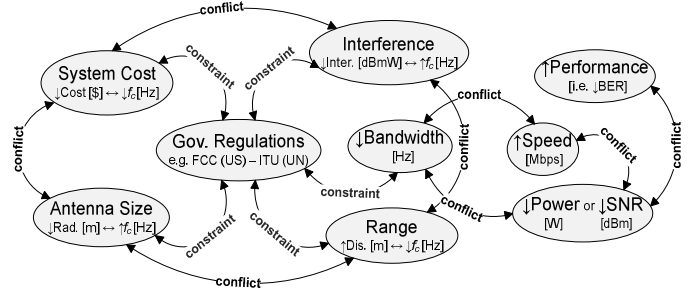

Fig. 2. Example of wireless system optimization problem.

In this example, perhaps the most important aspect of radio communications is the operating frequency and the assigned passband spectrum range. To a large extent, a license of operation is required from governmental bodies that oversee the frequency allocation (e.g. United States FCC and DoC, Industry Canada, United Nations ITU, etc.). The challenge here is the difficulty to obtain an unused or uncrowded band and the very high cost to get the license. Granted, some short range wireless technologies, such as: RFID, cordless phones, ZigBee, Wi-Fi, Bluetooth, and UWB can freely function in the unlicensed range commonly known as ISM and ultra-wide bands. On the other hand, mobile communications, and all other transmissions of the like, will require authorization and certification. As a consequence, this initial frequency assignment constraint creates a number of contradicting requirements; among them: radio interference, transmission range, system cost and complexity, antenna size, bandwidth capacity. And accordingly, because of further interdependencies, these elements will cause other disagreements with the link data rate, power requirements, performance, and so on. In fact, as more elements are considered within the system, the design problem becomes more realistically inclusive, but will demand greater expertise and methods to meticulous adjust and fine-tune the numerous restrictions that have proportionally grown.

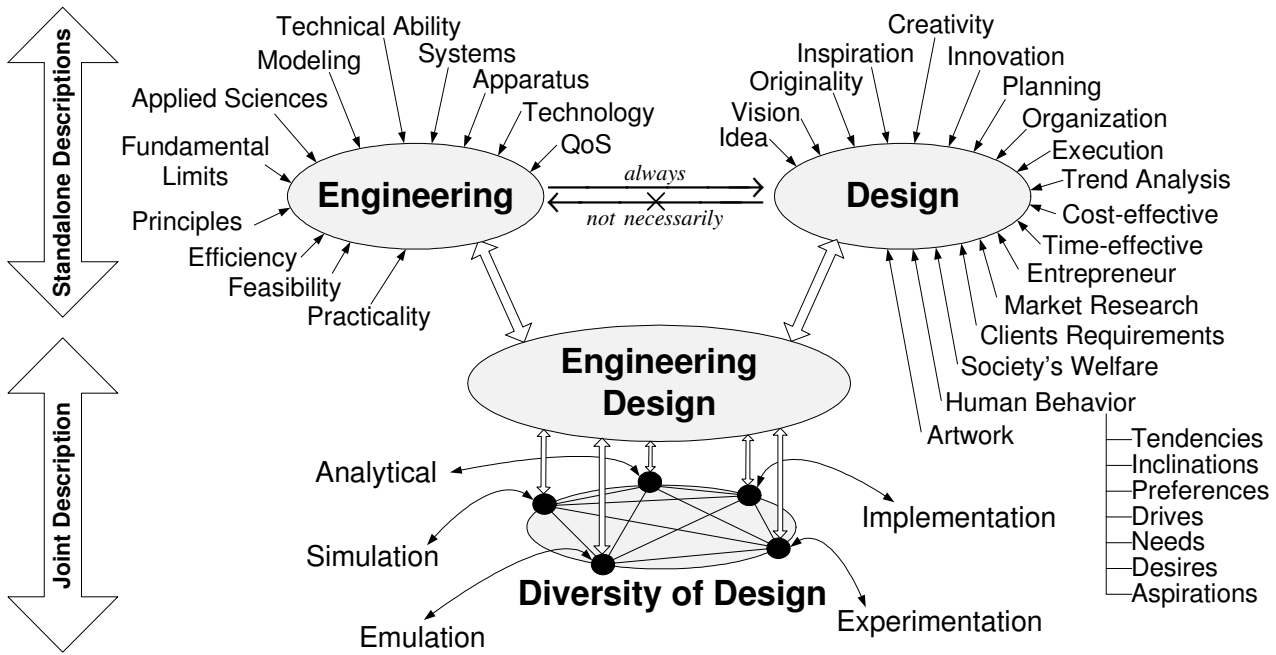

Fig. 1. A simple and coherent description of engineering design.

## 8. DIVERSITY OF DESIGN

As of yet, we have reflected on the notion of engineering design, and have expressed the paramount importance of design in engineering. At this level, it becomes interesting to proceed and briefly investigate the actual approach for performing design. Namely, how should engineers go about designing a sophisticated system such as the one explained above? And, what are the different mechanisms, tools, and strategies available for careful design?

Without a doubt, hands-on designing is necessary to better understand and optimize a particular system. But what exactly do we suggest by "hands-on"? From a first impression, it is safe to say that this term does implicate some sort of hardware-based experimentation. In fact, it could, and rightly so, be argued that design absolutely requires physical experimentation for a robust outcome. But we should be careful not to forget that design can actually encompass more than one approach.

Of course, it is certain that technology has become pivotal in engineering. In particular, the processing capability of computers has changed dramatically the way scientific investigation and research is performed. In fact, simulation has a number of benefits, above all: it is cost-effective; it enables simple parameter modifications; and for the most part the results are quickly computed. Indeed, even Richard Hamming, the renowned Bell Labs researcher, predicted, and later demonstrated, that 90% of experiments would be conducted on computers and the remaining in traditional labs [9]. On the other hand, the distinguished MIT physicist and engineer Hermann Haus considered lab-based experimentation as one of his "philosophy of life" [10]. Without becoming a paradox, both arguments are valid, and in their own rights correct; sometimes, they even complement each other.

Thus, design diversity is rather desirable because it allows different engineers, skillful in a particular environment, to work on the same problem using variant means. Therefore, for the sake of accurate engineering, we may stretch the typical understanding of applied design to multiple domains, such as: analytical, simulation, emulation, experimentation, and implementation. Indeed, these methods can be seen as the practical interface between engineering scientists and the exercise of design. Eventually, insights from each of these methods could mutually be shared and compared among them in order to converge to an optimal solution. Hence, in Fig. 1, we visually depict the interconnectedness of these techniques.

## 9. CONCLUSION

In this paper, we tried to gradually articulate our viewpoint and understanding of engineering design by disaggregating the topic into various descriptive spheres. Specifically, we discussed the management, creativity, and execution of design. We then analyzed the connotation and explored the distinctive attributes of engineering design via a separate and joint perspective. We also demonstrated through an example the complexity of engineering design. Lastly, we commented on the idea of design diversity, and enticed the combination of various technical practices in order to produce great design as opposed to good design.